\documentclass[12pt]{article}

\usepackage{amsmath}
\usepackage{graphicx}
\usepackage{appendix}
\usepackage{authblk}
\usepackage{txfonts} 

\makeatother

\begin{document}

\title{Time Evolution of the Spread of Diseases with a General Infectivity
Profile on a Complex Dynamic Network}

\author[1]{Bahman Davoudi}

\author[1,2,*]{Babak Pourbohloul}

\affil[1]{Division of Mathematical Modeling, University of British
Columbia Centre for Disease Control, Vancouver, British Columbia,
Canada} 
\affil[2]{School of Population \& Public Health, University
of British Columbia, Vancouver, British Columbia, Canada} 
\affil[*]{Corresponding author} 

\setlength{\baselineskip}{1.5\baselineskip} 
\maketitle

\begin{abstract}
This manuscript introduces a new analytical approach for studying
the time evolution of disease spread on a finite size network. Our
methodology can accommodate any disease with a general infectivity
profile. This new approach is able to incorporate the impact of a
general intervention - at the population level - in a number of different
ways. Below, we discuss the details of the equations involved and
compare the outcomes of analytical calculation against simulation
results. We conclude with a discussion of possible extensions of this
methodology. 
\end{abstract}

\section{Introduction}

The time evolution of disease spread within human populations is a
very interesting and multifaceted topic; it is closely related to
the rate of new infections in a population, which has significant
implications when designing public health policy. The spread of disease
within a population is complex and depends on a number of different
factors, including, social connectivity patterns, cultural practices
and education surrounding hygiene and intervention strategies, the
level of preexisting immunity and finally, the infectivity of the
infectious agent. In order to produce a reliable estimate of the new
infection rate, a model should incorporate - at the very least - the
abovementioned factors. 

The spread of disease in a population involves a complex stochastic
branching process. This process is comprised of three distinct constituents,
namely, the stochastic phase, exponential phase and declining phase.
To date, each phase has undergone a considerable amount of scrutiny
using two main techniques, i.e., compartmental and network models.
The stochastic phase was studied using discrete and continuous time
approaches \cite{Marder_2007,Noel_2008,Davoudi_2009,Miller_2009},
while the exponential and declining phases were studied using a variety
of models and techniques, including, compartmental \cite{Bailey_1975,Anderson_1991,Brauer_2008}
and network models \cite{Noel_2008,Davoudi_2009,Miller_2009,Volz_2007,Davoudi_2010}.
Both of these method, however, have their downfalls; that is, compartmental
models deal solely with constant infection and removal rates, do not
incorporate any memory of infections, distribute the force of infection
uniformly in the population, and include the finite size effect in
a very simple manner. Moreover, some of the limitations associated
with network methods are seen in the work by Nel \textit{et.al} \cite{Noel_2008}
and Davoudi\textit{ et.al} \cite{Davoudi_2010}. They considered disease
transmission to follow the generation time concept, that is, they
assumed infection or recovery/removal occurred at a discrete time
within the period of $\tau$. This led to a significant simplification
of the calculations, thus, the model became unrealistic for a disease
with a long period of infection. The work by Volz\textit{ et.al} \cite{Volz_2007}
included the finite size effect in a comprehensive manner, but the
method could not accommodate diseases with a complex infection profile.
To circumvent these downfalls we present a new methodology which can
sustain complex infection and recovery/removal profiles, and can take
the finite size effect into account, in a precise manner.

Below, we first discuss the required theoretical components for this
methodology. We then compare theoretical results against simulation
results to understand the precision of the method and the level of
approximation involved. Subsequently, we discuss the possibility of
extending this current methodology to address systems with a complex
dynamic network structure.

\section{Theory}

\subsection{Basic Notions}

We consider the preexisting contact network to be the platform for
disease transmission within a population. Individuals are represented
by a vertex and contact between two individuals is represented by
a link. Links are composed of two stubs, each of which is attached
to its own vertex. For a population of size $N$, we create a random
network where the degree of connectivity between vertices is coded
by the probability distribution function, $p_{k}$, where $k$ is
the degree. The average excess degree\cite{Noel_2008} is defined
as the degree of secondarily infected vertices and is given by $Z_{x}=z_{2}/z_{1}$
where $z_{1}=\langle k\rangle_{p_{k}}$ is the average degree, $z_{2}=\langle k^{2}\rangle_{p_{k}}-\langle k\rangle_{p_{k}}$
and finally $\langle k^{n}\rangle_{p_{k}}=\sum_{\kappa=0}^{\infty}k^{n}p_{\kappa}$.

The process of disease transmission within a population is a complex
phenomenon. Individuals, or vertices, are infected and removed at
specific times, denoted by $t_{i}$ and $t_{r}$ respectively. Infected
vertices are removed from the population for a number of reasons,
including death, quarantine, and recovery for example. Moreover, the
age of infection and the time between infection and removal, or the
{}``time-to-removal'',for each infected vertex are defined as $\tau=t-t_{i}$
and $\tau_{r}=t_{r}-t_{i}$, respectively. The transmission of infection
between an infected vertex and a susceptible vertex is dictated by
the infectivity function, $\lambda_{i}(\tau)$, in which $\lambda_{i}(\tau)d\tau$
denotes the probability of transmission during the interval $\tau$
and $\tau+d\tau$. In the same manner, we define the removal function
as $\lambda_{r}(\tau)$, in which $\lambda_{r}(\tau)d\tau$ yields
the chance of removal of an infected vertex during the interval $\tau$
and $\tau+d\tau$. The transmissibility, $T(\tau,\tau_{r})$, provides
the probability of transmission until time $\tau$, for an infected
vertex that is removed after time $\tau_{r}$ and satisfies the following
equation\cite{Cox_1984,Newman_2002} \begin{equation}
T(\tau,\tau_{r})=\left\{ \begin{array}{ccc}
1-\exp\left(-\int_{0}^{\tau}\lambda_{i}(u)du\right) &  & \tau<\tau_{r}\\
\\1-\exp\left(-\int_{0}^{\tau_{R}}\lambda_{i}(u)du\right)\;\; &  & {otherwise}\end{array}\right.\label{eq:Ttr}\end{equation}

We define $\Psi(\tau_{r})$\cite{Cox_1984} as the probability that
a vertex has a time-to-removal $\ge\tau_{r}$, which is given by \begin{equation}
\Psi(\tau_{r})=\exp\left(-\int_{0}^{\tau_{R}}\lambda_{r}(u)du\right),\label{eq:Psi}\end{equation}
 subject to the condition $\Psi(\infty)=0$. We also define the probability
density function as $\psi(\tau_{r})=-\frac{d\Psi(\tau_{r})}{d\tau_{r}}$
(or $\Psi(\tau_{r})=\int_{\tau_{r}}^{\infty}\psi(u)du$). The basic
reproduction number is given by ${\cal R}_{0}=\frac{z_{2}}{z_{1}}T$,
where $T$ is the ultimate transmissibility $T=\int_{0}^{\infty}\psi(\tau)T(\tau,\tau<\tau_{r})$\cite{Davoudi_2009}.

Finally, $J(t)$ denotes the rate of newly infected vertices at time
$t$. At a given time $t$, a fraction, $\Psi(\tau)$, of infected
vertices $J(t-\tau)$ with the age of infection $\tau$ remain infectious.
Therefore, the total number of infectious vertices, at a given time,
can be calculated by \begin{equation}
N_{i}(t)=\int_{0}^{t}J(t-\tau)\Psi(\tau)d\tau\label{eq:Ni}\end{equation}
 and the number of removed and susceptible vertices are given by \begin{equation}
N_{r}(t)=\int_{0}^{t}J(t-\tau)\left[1-\Psi(\tau)\right]d\tau,\label{eq:Nr}\end{equation}

and\begin{equation}
N_{s}(t)=N-N_{r}(t)-N_{i}(t).\label{eq:Ns}\end{equation}

respectively.

\subsection{Disease Transmission Dynamics on a Network}

In the following section we introduce and discuss the set of equations
we use to find the rate of new infections $J(t)$ as a function of
time. This calculation becomes possible once we combine the network
aspects (vertices connectivity) with the disease status of each vertex.
To elaborate, we start with one infectious vertex with excess degree
$Z_{x}$, and assume that it was infected at time $t_{i}=0$. With
this knowledge, we then calculate the number of new infection that
arose from this infected vertex, at the later time $t=\tau$, using
\begin{equation}
J(\tau)=\Psi(\tau)\frac{dT(\tau,\tau<\tau_{r})}{d\tau}Z_{x},\label{eq:dJ}\end{equation}
 where $\frac{dT(\tau,\tau<\tau_{r})}{d\tau}$ is the contribution
of each link to disease transmission between time $\tau$ and $\tau+d\tau$,
given that the vertex was not removed by time $t$; the resulting
contribution of $Z_{x}$ link is given by $\frac{dT(\tau,\tau<\tau_{r})}{d\tau}Z_{x}$.
The equation above is then multiplied by $\Psi(\tau)$ in order to
take the chance of removal into account. The total number of infections
caused by the first infected vertex is given by $TZ_{x}$ where $T=\int_{0}^{\infty}\Psi(\tau)\frac{dT(\tau,\tau<\tau_{r})}{d\tau}d\tau=\int_{0}^{\infty}\psi(\tau)T(\tau,\tau<\tau_{r})d\tau$
is the ultimate transmissibility, which yields the probability of
infection along a link.

Equation \eqref{eq:dJ} can be easily extended to the initial phase
of an epidemic, assuming that the excess degree of all vertices is
the same. In general, the renewal equation for $J(t)$ is as follows\cite{Lotka_1939,Davoudi_2009}
\begin{equation}
J(t)=\int_{0}^{t}J(t-\tau)\Psi(\tau)\frac{dT(\tau,\tau<\tau_{r})}{d\tau}Z_{x}(t,\tau)d\tau\label{eq:Jt}\end{equation}
 The right hand side of the above equation gives the total number
of transmitting links $\lambda(t)$ at time $t$, which leads to $J(t)\approx\lambda(t)$
infections\cite{Davoudi_2010}. Equation \eqref{eq:Jt}, where $Z_{x}(t,\tau)\approx Z_{x}=z_{2}/z_{1}$,
can be used before the finite size effect becomes important, which
is a valid assumption while $N_{r}(t)+N_{i}(t)\ll N_{s}(t)$. In the
limit $N_{r}(t)+N_{i}(t)\sim N_{s}(t)$ , only a fraction of $Z_{x}$
is used to connect the infected and susceptible vertices. An appropriate
approximation for $Z_{x}(t,\tau)$ is given below.

$Z_{x}(t,\tau)$ is calculated in two steps. First, the typical degree
of infected vertices is calculated during the process of disease transmission.
Second, an estimate of the average number of links an infected vertex
- with an infectious period of $\tau$ - could have with susceptible
vertices, at time $t$, is made.

The first step is easily preformed for a random network \cite{Noel_2008,Davoudi_2010}.
A vertex is randomly picked and assigned to a collected class, while
the probability that the chosen vertex has the degree $k$ is given
by $q_{k}(1)=p_{k}$. The function argument shows the number of collected
vertices. The degree distribution of the uncollected vertices is given
by $p_{k}(1)=p_{k}$. The expected degree of the first collected vertex
is $\tilde{z}(1)=z_{1}$ and the average degree of the uncollected
vertices is $z(1)=z_{1}$. A second vertex is then chosen by picking
a random stub. The probability that the second vertex has degree $k$
is given by $q_{k}(2)=kp_{k}/z(1)$, the degree distribution of the
collected vertices is calculated as follows $\tilde{p}_{k}(2)=(q_{k}(1)+q_{k}(2))/2$
and the degree distribution of the uncollected vertices is specified
by $p_{k}(2)=(Np_{k}-2\tilde{p}_{k}(2))/(N-2)$. The probability that
the $nth$ chosen vertex has degree $k$ is $q_{k}(n)=kp_{k}(n-1)/z(n-1)$
and, in the same manner, the degree distributions of collected and
uncollected vertices are given by \begin{align}
\tilde{p}_{k}(n) & =\sum_{i=1}^{n}q_{k}(n)/n,\label{eq:pk}\\
p_{k}(n) & =(Np_{k}-n\tilde{p}_{k}(n))/(N-n),\label{eq:pkt}\end{align}
 where $z(n)=\sum_{i=0}^{\infty}kp_{k}(n)$ is the average degree
of uncollected vertices after $n$ collections. We define $\tilde{z}(n)=\sum_{i=0}^{\infty}k\tilde{p}_{k}(n)$
as the average degree of collected vertices after $n$ collections.
The latter equations take the following form in the continuous limit
\begin{align}
\frac{d\tilde{p}_{k}(n)}{dn} & =\frac{p_{k}(n)}{n}\left(\frac{k}{z(n)}-1\right),\label{eq:pktc}\\
\frac{dp_{k}(n)}{dn} & =\frac{p_{k}(n)}{N-n}\left(1-\frac{k}{z(n)}\right).\label{eq:pkc}\end{align}
 The average degree of infected vertices that are infected between
time $t$ and $t+dt$ is given by \begin{align}
z_{j}(t) & =\frac{[N_{r}(t)+N_{i}(t)+J(t)dt]\tilde{z}(N_{r}(t)+N_{i}(t)+J(t)dt)-(N_{r}(t)+N_{i}(t))\tilde{z}(N_{r}(t)+N_{i}(t))}{J(t)dt}\\
 & =\tilde{z}(N_{r}(t)+N_{i}(t))+[N_{r}(t)+N_{i}(t)]\left.\frac{d\tilde{z}(n)}{dn}\right|_{n=N_{r}(t)+N_{i}(t)}.\label{eq:zj}\end{align}
 In the same manner, we find that the average degree of removed $z_{r}(t)$,
infectious $z_{i}(t)$, and susceptible classes $z_{s}(t)$ with the
following formulas, respectively \begin{align}
z_{r}(t) & =\tilde{z}(N_{r}(t)),\label{eq:zr}\\
z_{i}(t) & =\frac{[N_{r}(t)+N_{i}(t)]\tilde{z}(N_{r}(t)+N_{i}(t))-N_{r}(t)\tilde{z}(N_{r}(t))}{N_{i}(t)},\label{eq:zi}\\
z_{s}(t) & =z(N_{s}(t)),\label{eq:zs}\end{align}

$Z_{x}(t,\tau)$ can now be easily estimated. The number of stubs
belonging to infected vertices, between time $t-\tau$ and $t-\tau+d\tau$,
is given by $z_{j}(t-\tau)J(t-\tau)d\tau$. The probability of one
of these stubs connecting to the stub of a susceptible vertex, at
time $t$, is given by $z_{s}(t)N_{s}(t)/z_{1}N$, and thus \begin{equation}
Z_{x}(t,\tau)\approx z_{j}(t-\tau)\frac{z_{s}(t)N_{s}(t)}{z_{1}N}.\end{equation}

We also considered other approximation with different renewal equations
\begin{equation}
J(t)=\int_{0}^{t}J(t-\tau)\Psi(\tau)\frac{dT(\tau,\tau<\tau_{r})}{d\tau}Z_{x}N_{s}(t)/Nd\tau,\label{eq:JComp}\end{equation}
 This new methodology provides a solution for SIR compartmental models,
whereby the force of infection, $\beta(\tau)$, and recovery rate,
$\gamma(\tau)$, are functions of $\tau$ \cite{Miller_2009}. This
is equivalent to the approximation $Z_{x}(t,\tau)\approx Z_{x}N_{s}(t)/N$.

\section{Numerical results for the most simplistic network}

\label{sec:Num_simp_net} 
\begin{figure}[htb]
\begin{centering}
\includegraphics[scale=0.5]{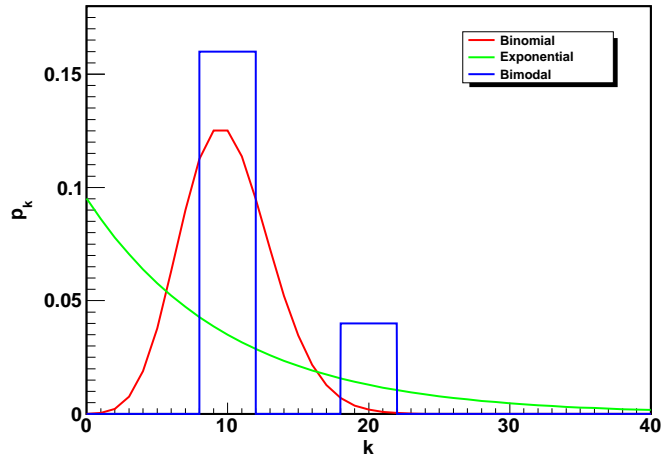}
\caption{Three different degree distributions, binomial $z_{1}=10$, exponential
$\kappa=10$ and bimodal network.\label{fig:pk}}
\end{centering}
\end{figure}

In this section we present the numerical
result for three stylized networks, namely binomial $p_{k}=\binom{N}{k}p^{k}(1-p)^{N-k}$
with $p=z_{1}/(N-1)$ ($z_{1}=10$), exponential $p_{k}=(1-\exp(-1/\kappa))\exp(-k/\kappa)$
($\kappa=10$) and bimodal, which are depicted in figure \ref{fig:pk}.
The size of the networks were $N=10,000$. To obtain a numerical solution
with the current methodology, we first created two sequences, namely
$z(n)$ and $\tilde{z}(n)$, for a specific network using equation
\eqref{eq:pk} and \eqref{eq:pkt}. We then recursively used the renewal
equation \eqref{eq:Jt} to obtain the number of infections at a later
time, meanwhile $z_{j}(t)$, $z_{s}(t)$ and $Z_{x}(t,\tau)$ were
calculated. %

\begin{figure}[htb]
\begin{centering}
\includegraphics[scale=0.35]{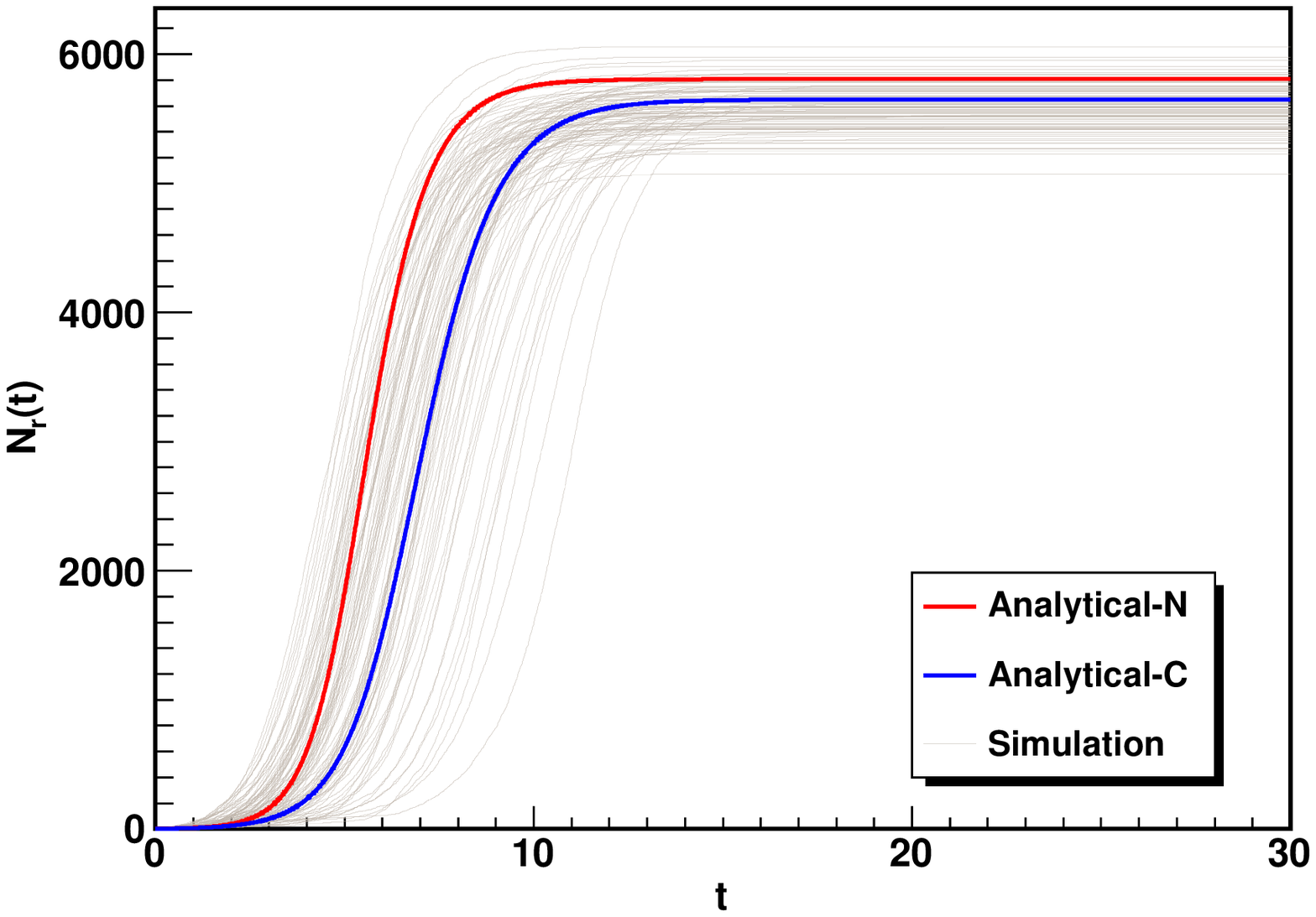}
\includegraphics[scale=0.35]{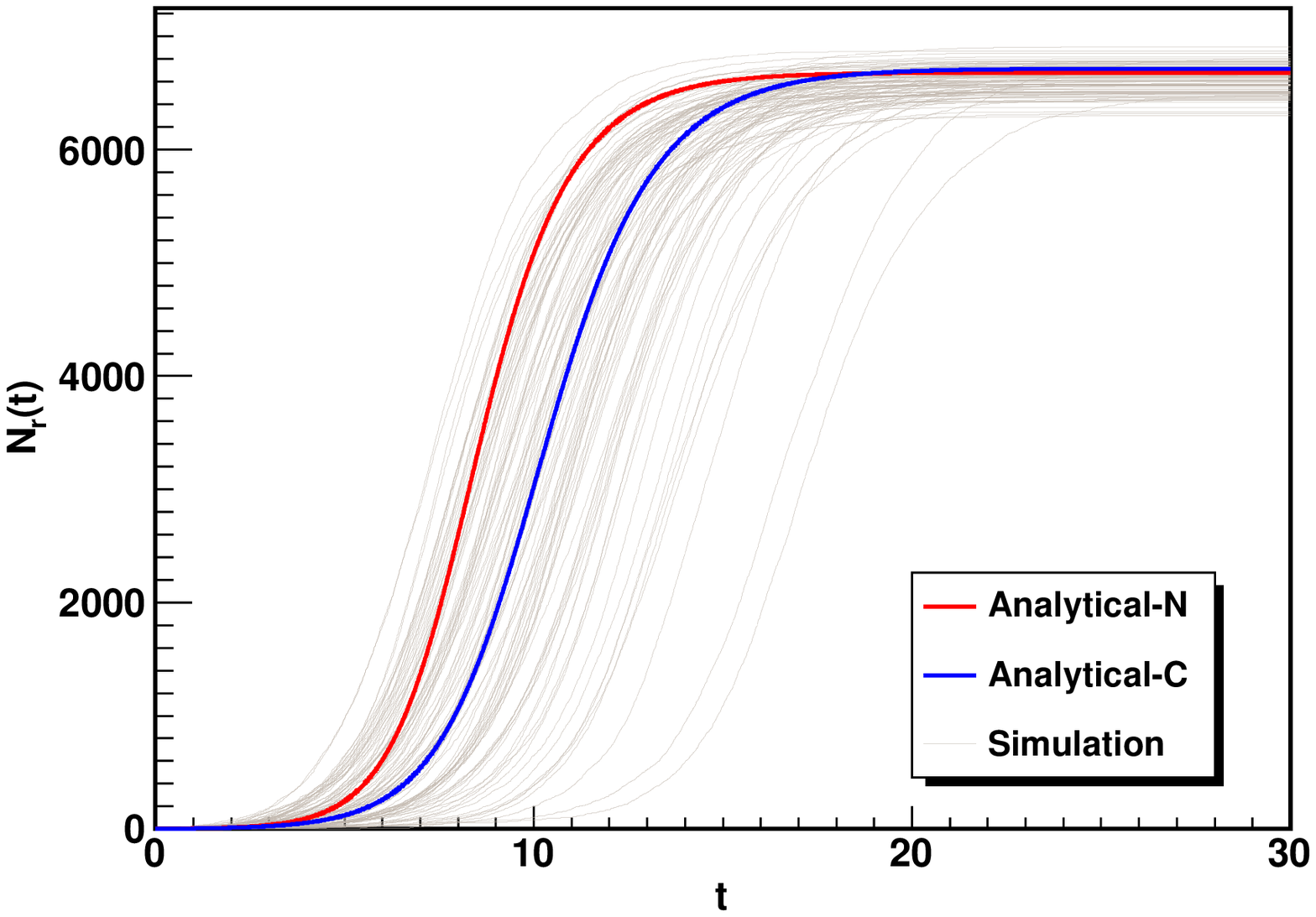}
\includegraphics[scale=0.35]{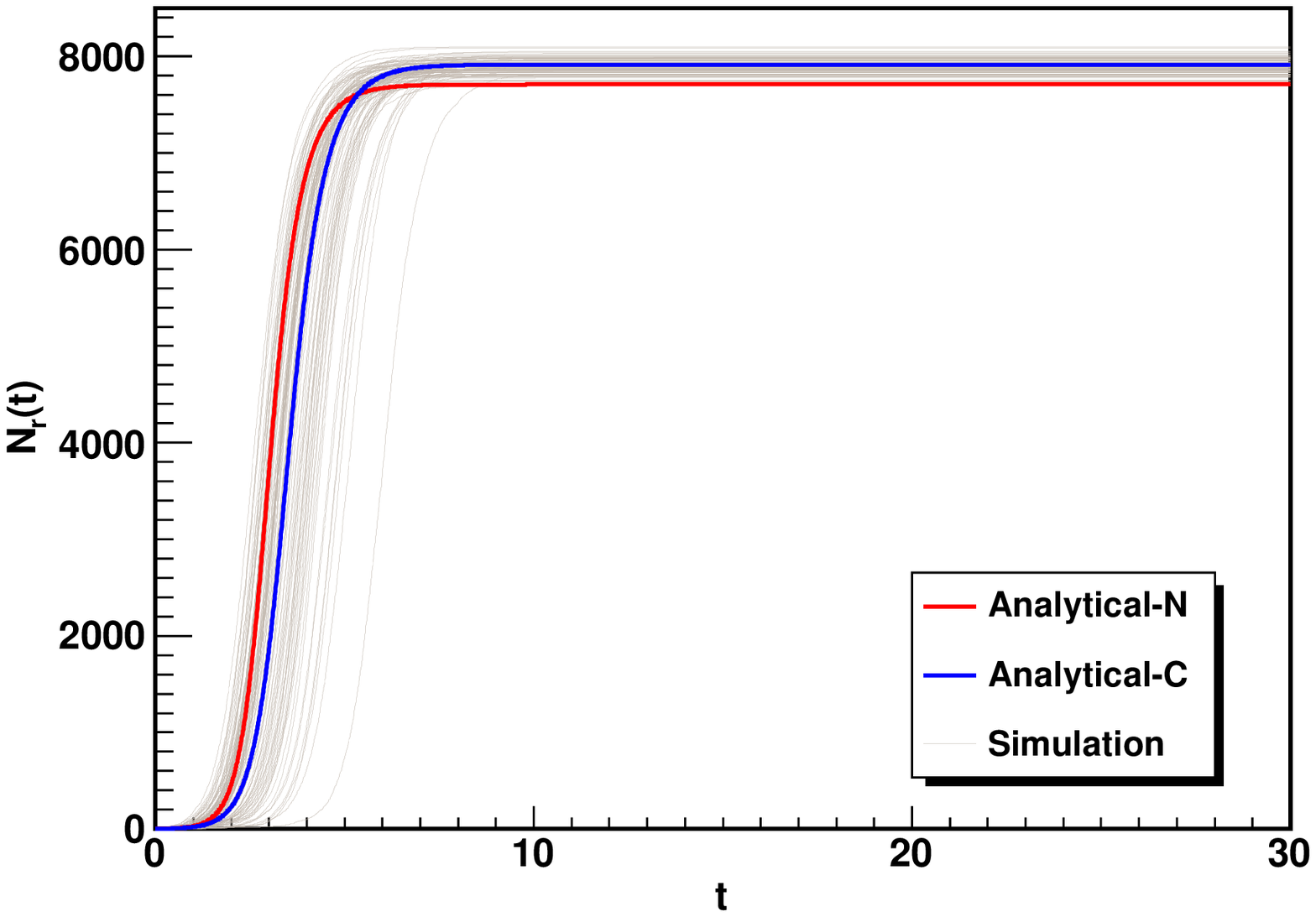}
\includegraphics[scale=0.35]{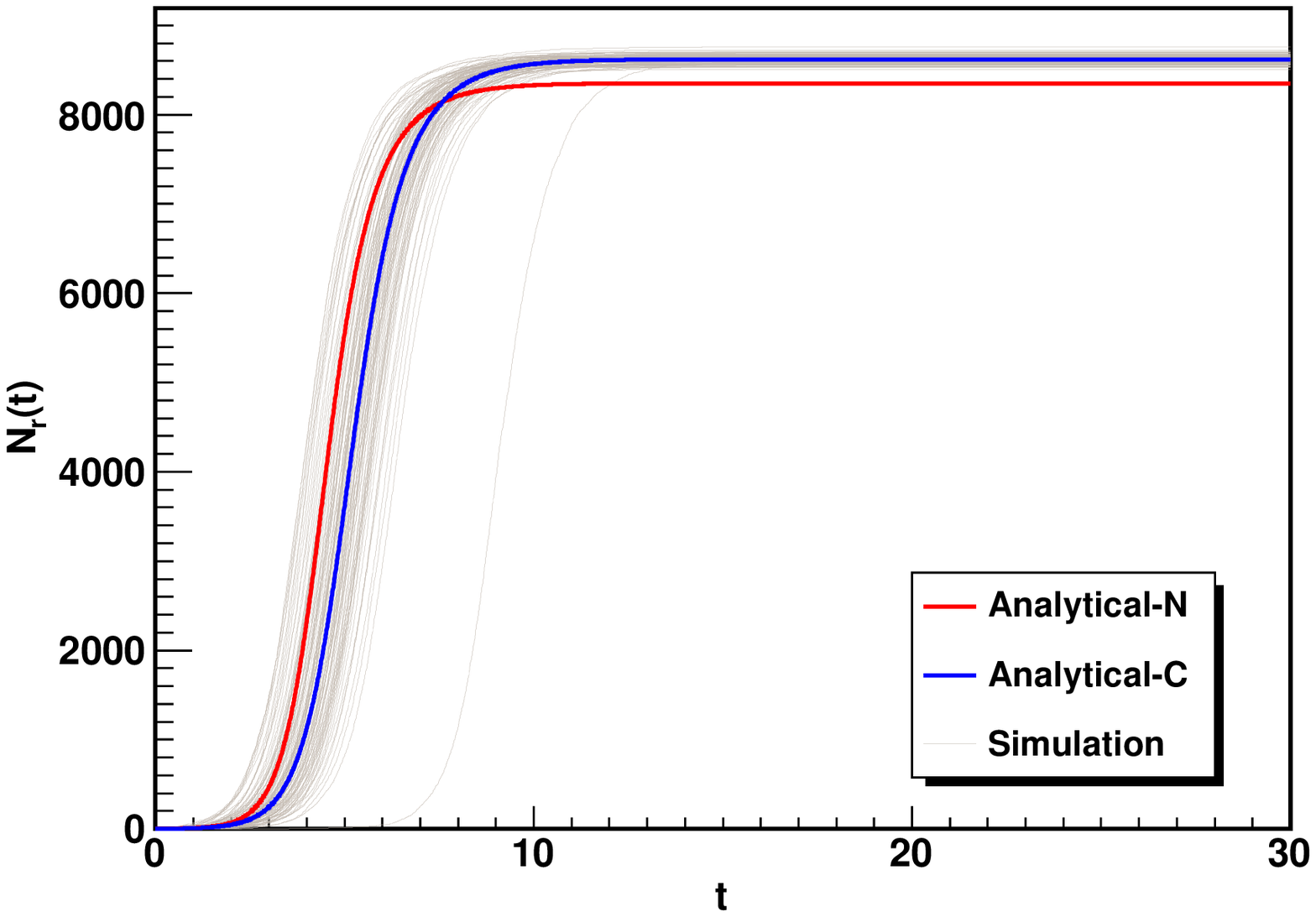}
\caption{$z_{1}=10$, $\lambda_{i}=0.25$ and $\lambda_{r}=2$ left top panel,
$\lambda_{i}=0.2$ and $\lambda_{r}=1$ right top panel, $\lambda_{i}=0.5$
and $\lambda_{r}=2$ left bottom panel, and finally $\lambda_{i}=0.3$
and $\lambda_{r}=1$ right bottom panel.\label{fig:fig1}}
\end{centering}
\end{figure}

In figure \ref{fig:fig1}, we compare the number of removed vertices
for the new calculation (Analytical-N) and the compartmental model
(Analytical-C), against simulations for a binomial network. Top left
panel $z_{1}=10$, $\lambda_{i}=0.35$ and $\lambda_{r}=2$; top right
panel $\lambda_{i}=0.2$ and $\lambda_{r}=1$; bottom left panel $\lambda_{i}=0.5$
and $\lambda_{r}=2$; and bottom right panel $\lambda_{i}=0.3$ and
$\lambda_{r}=1$. Both analytical approaches performed well for the
binomial network, within this range of parameter values. The current
approach slightly overestimated the final size for small ${\cal R}_{0}$
values and slightly underestimated the final size for large ${\cal R}_{0}$
values.

\begin{figure}[htb]
\begin{centering}
\includegraphics[scale=0.35]{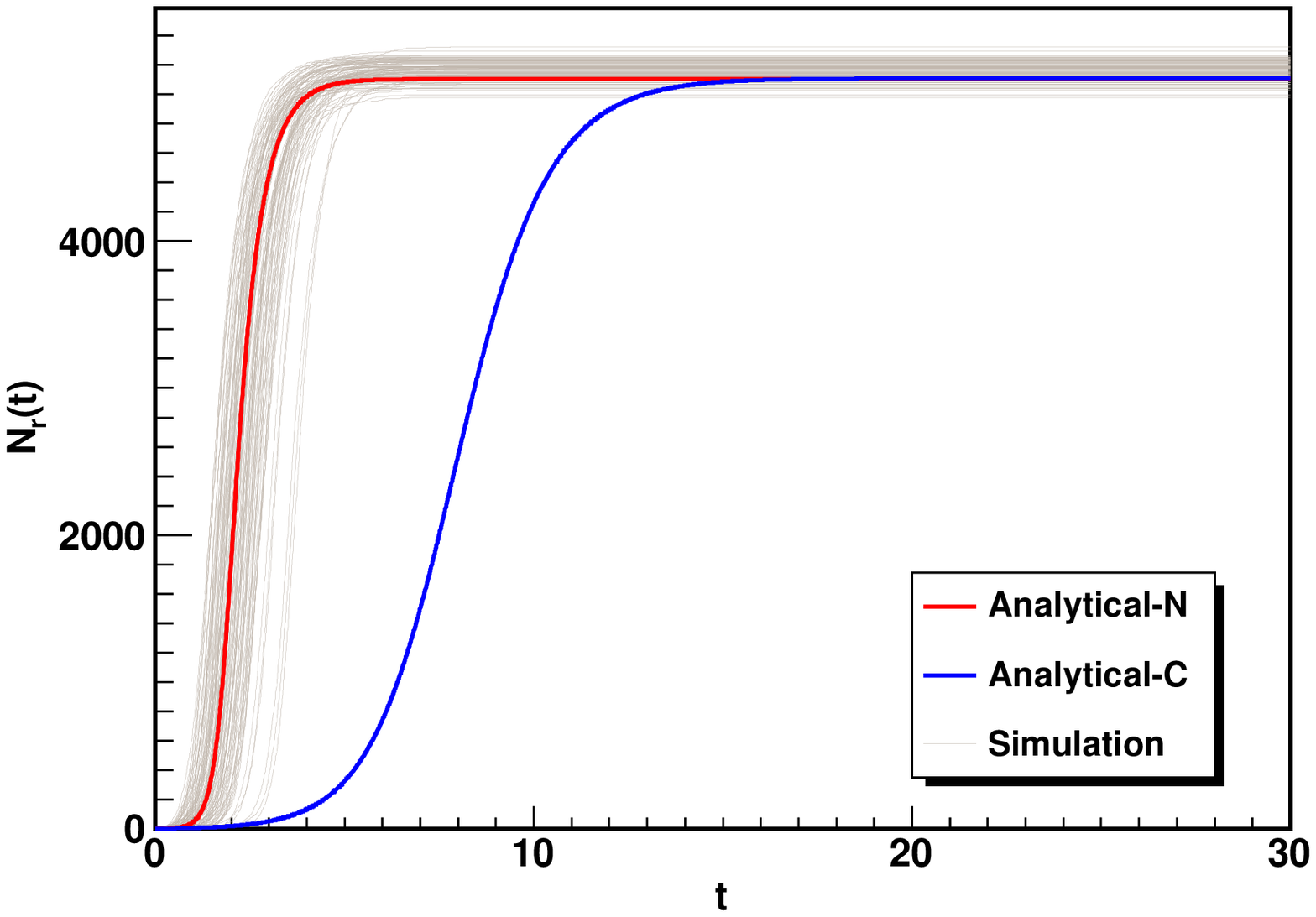}
\includegraphics[scale=0.35]{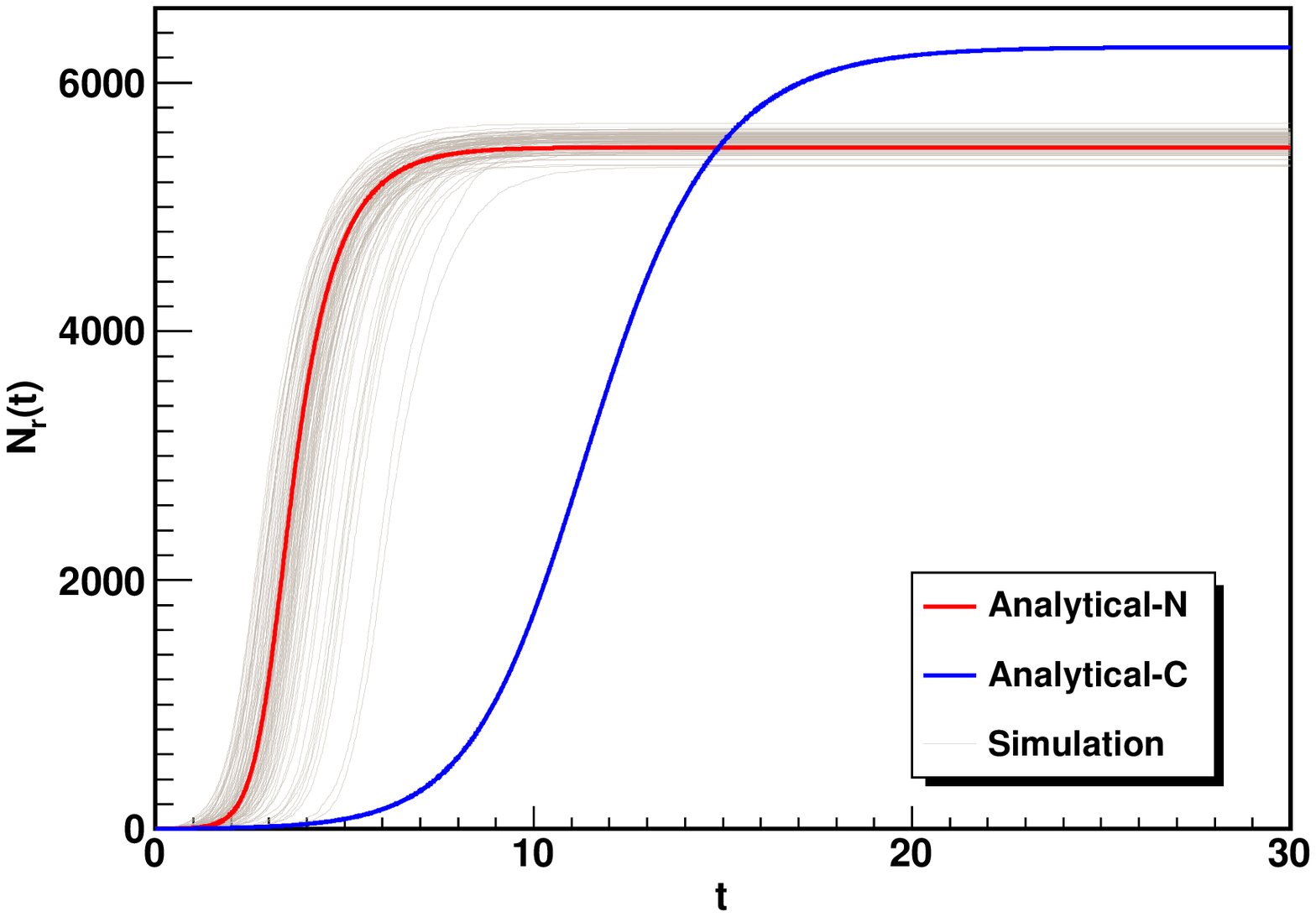}
\includegraphics[scale=0.35]{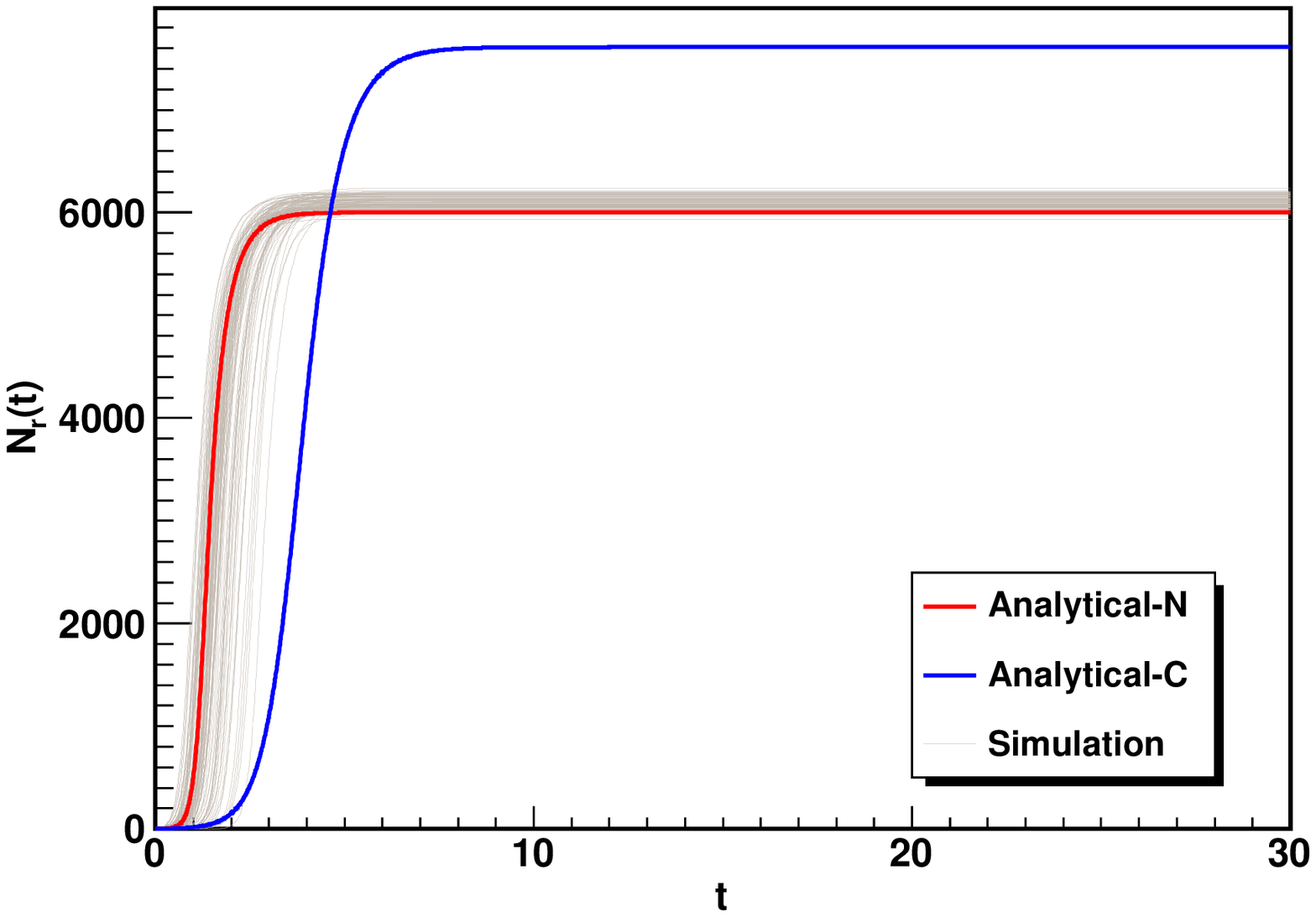}
\includegraphics[scale=0.35]{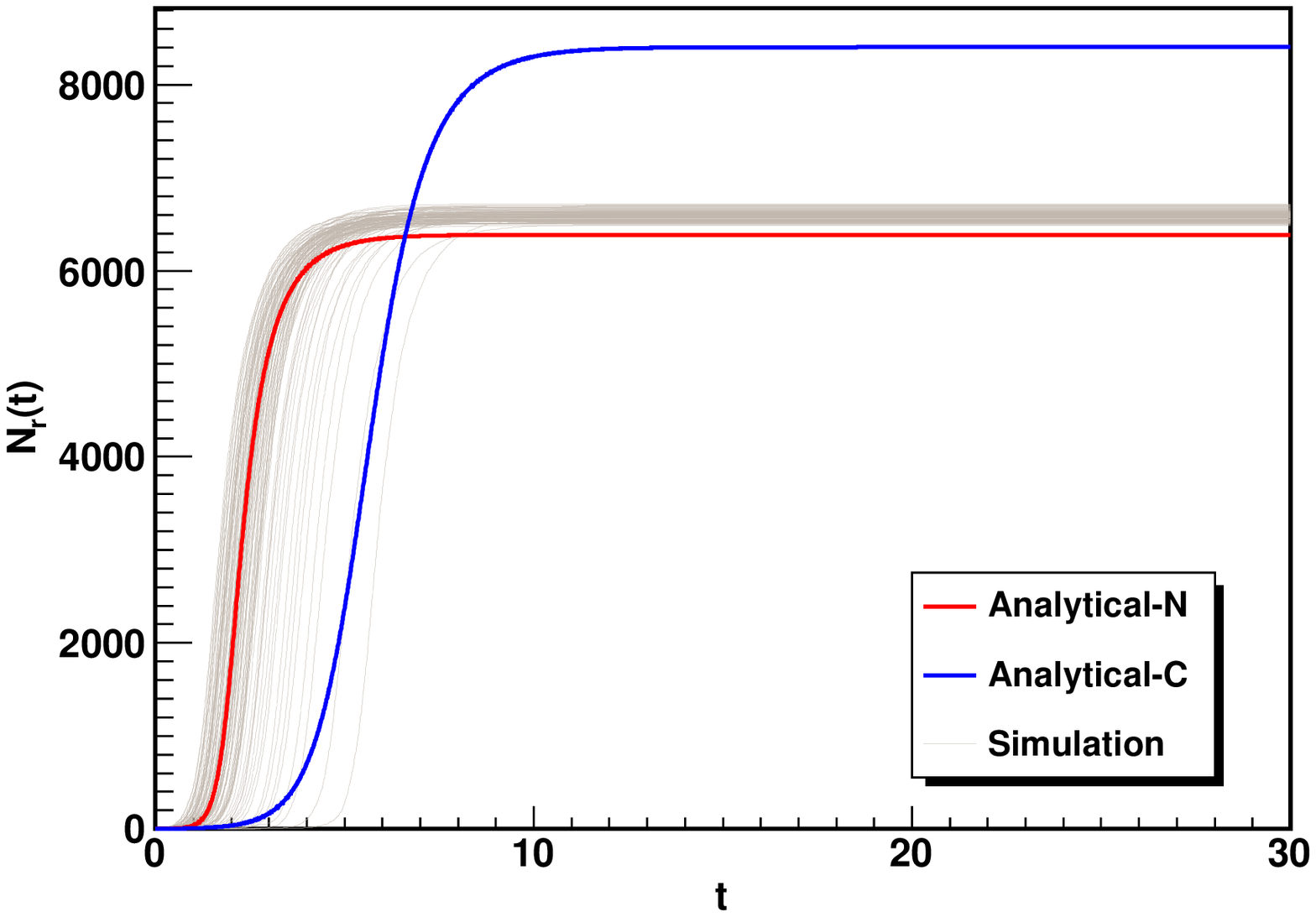}
\caption{$\kappa=10$, $\lambda_{i}=0.25$ and $\lambda_{r}=2$ left top panel,
$\lambda_{i}=0.2$ and $\lambda_{r}=1$ right top panel, $\lambda_{i}=0.5$
and $\lambda_{r}=2$ left bottom panel, and finally $\lambda_{i}=0.3$
and $\lambda_{r}=1$ right bottom panel.\label{fig:fig2}}
\end{centering}
\end{figure}

In figure \ref{fig:fig2}, we compare the number of removed vertices
for the new calculation (Analytical-N) and compartmental model (Analytical-C),
against simulations for exponential network. Top left panel $\kappa=10$,
$\lambda_{i}=0.35$ and $\lambda_{r}=2$; top right panel $\lambda_{i}=0.2$
and $\lambda_{r}=1$; bottom left panel $\lambda_{i}=0.5$ and $\lambda_{r}=2$;
and, bottom right panel $\lambda_{i}=0.3$ and $\lambda_{r}=1$. The
compartmental model was inferior to thebinomial network; for exponential
networks, as we expect that $Z_{x}(t,\tau)\approx Z_{x}$ will become
a poor predictor for any network with a very wide degree distribution.

\begin{figure}[htb]
\begin{centering}
\centering{}
\includegraphics[scale=0.35]{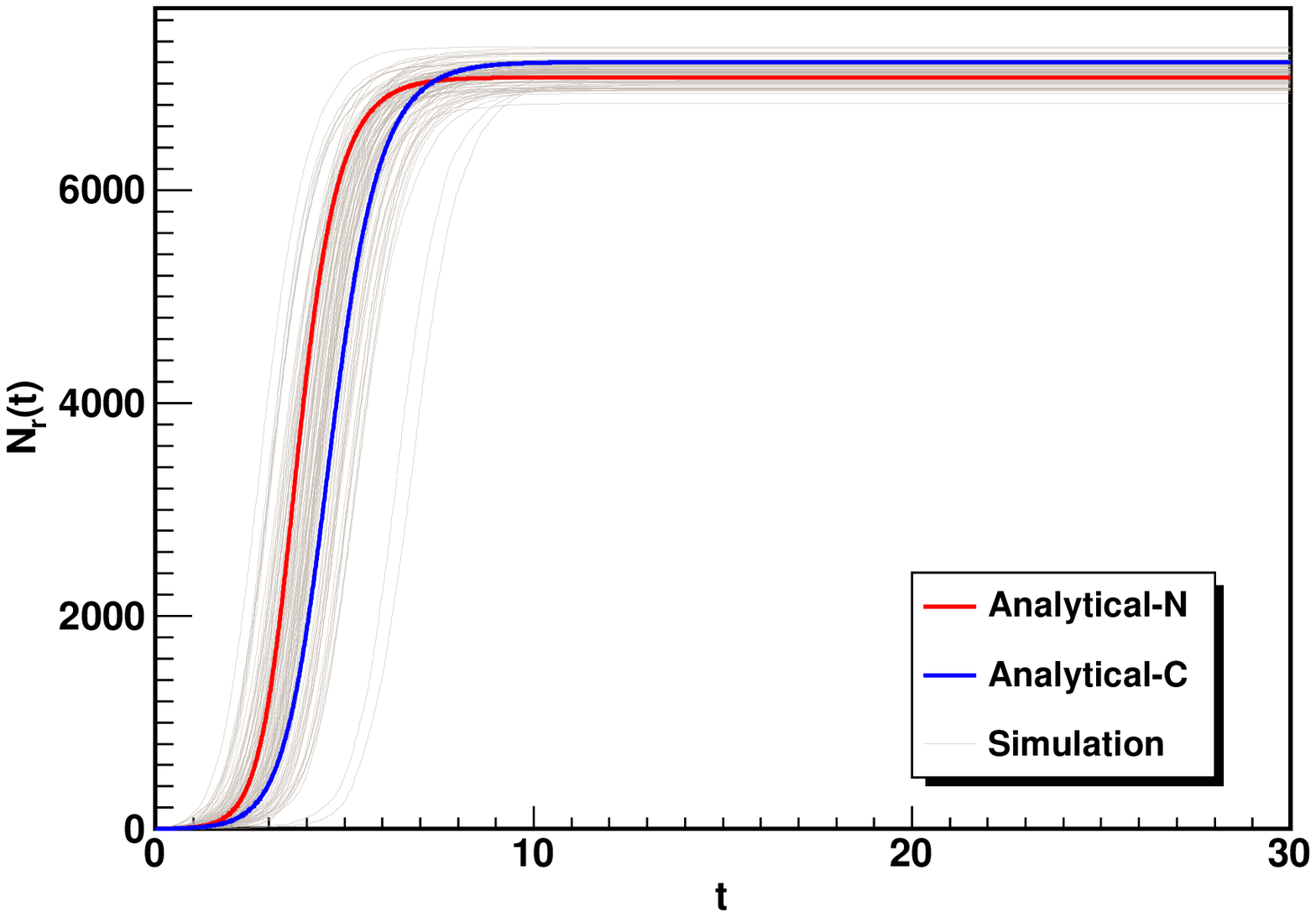}
\includegraphics[scale=0.35]{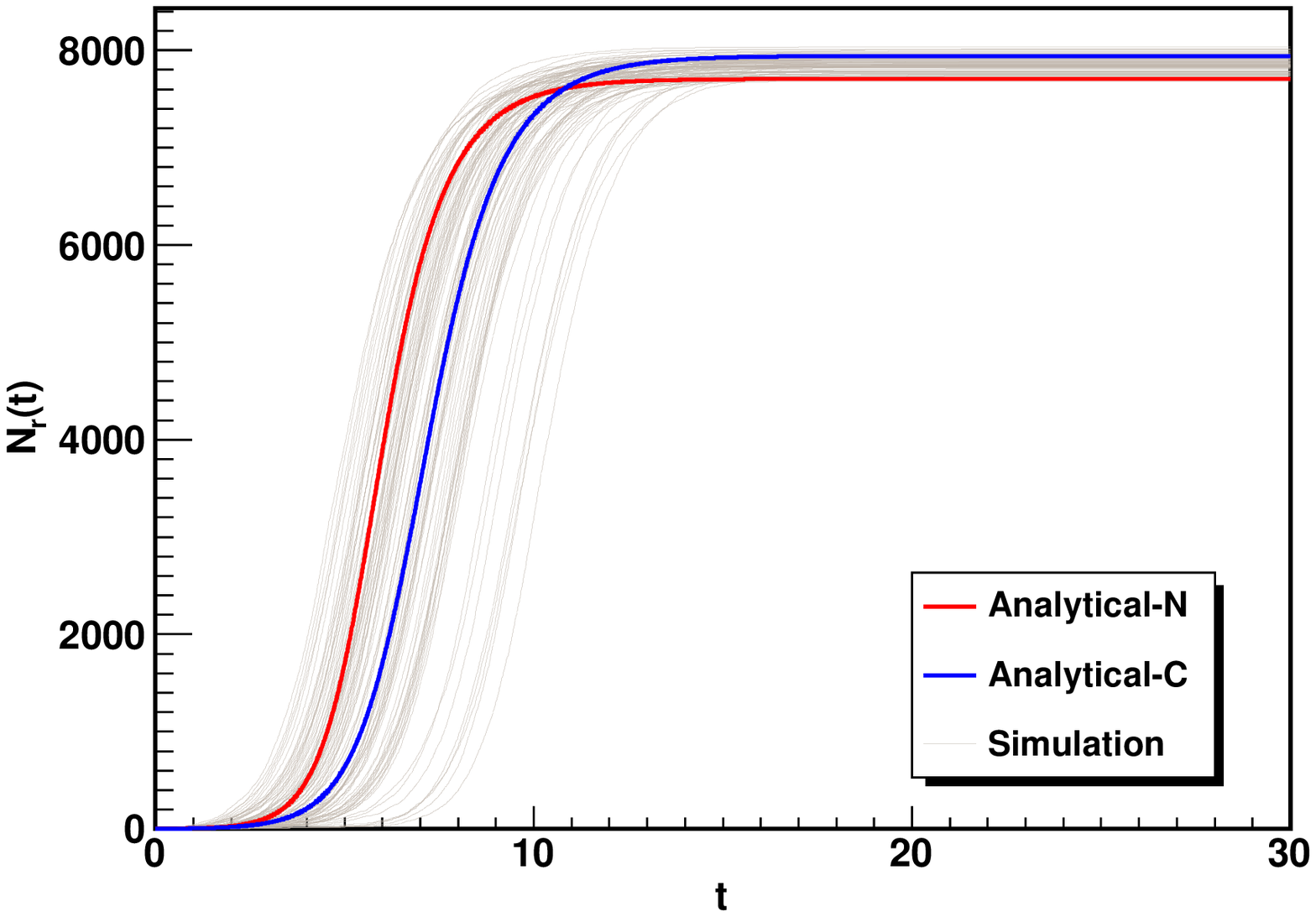}
\includegraphics[scale=0.35]{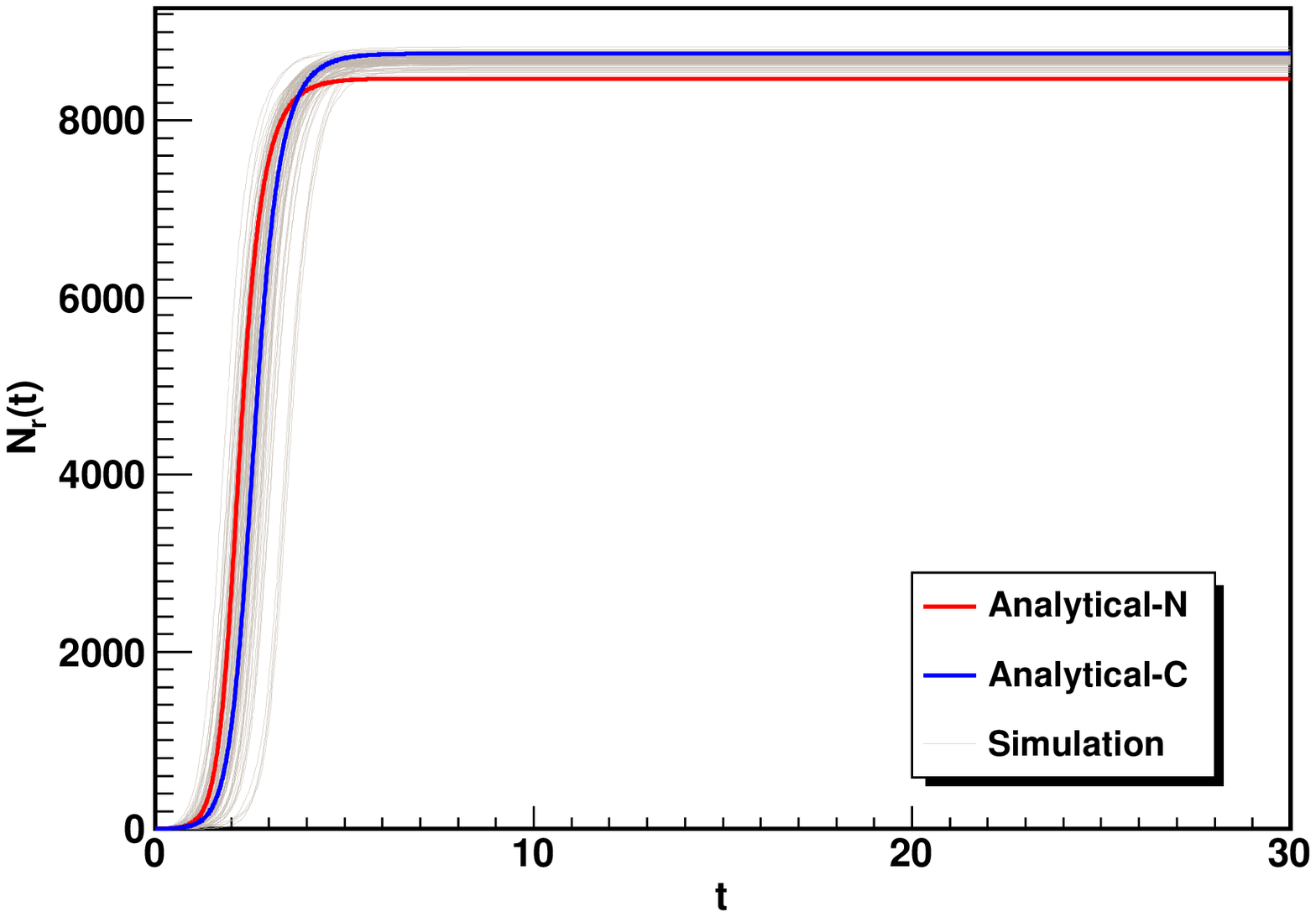}
\includegraphics[scale=0.35]{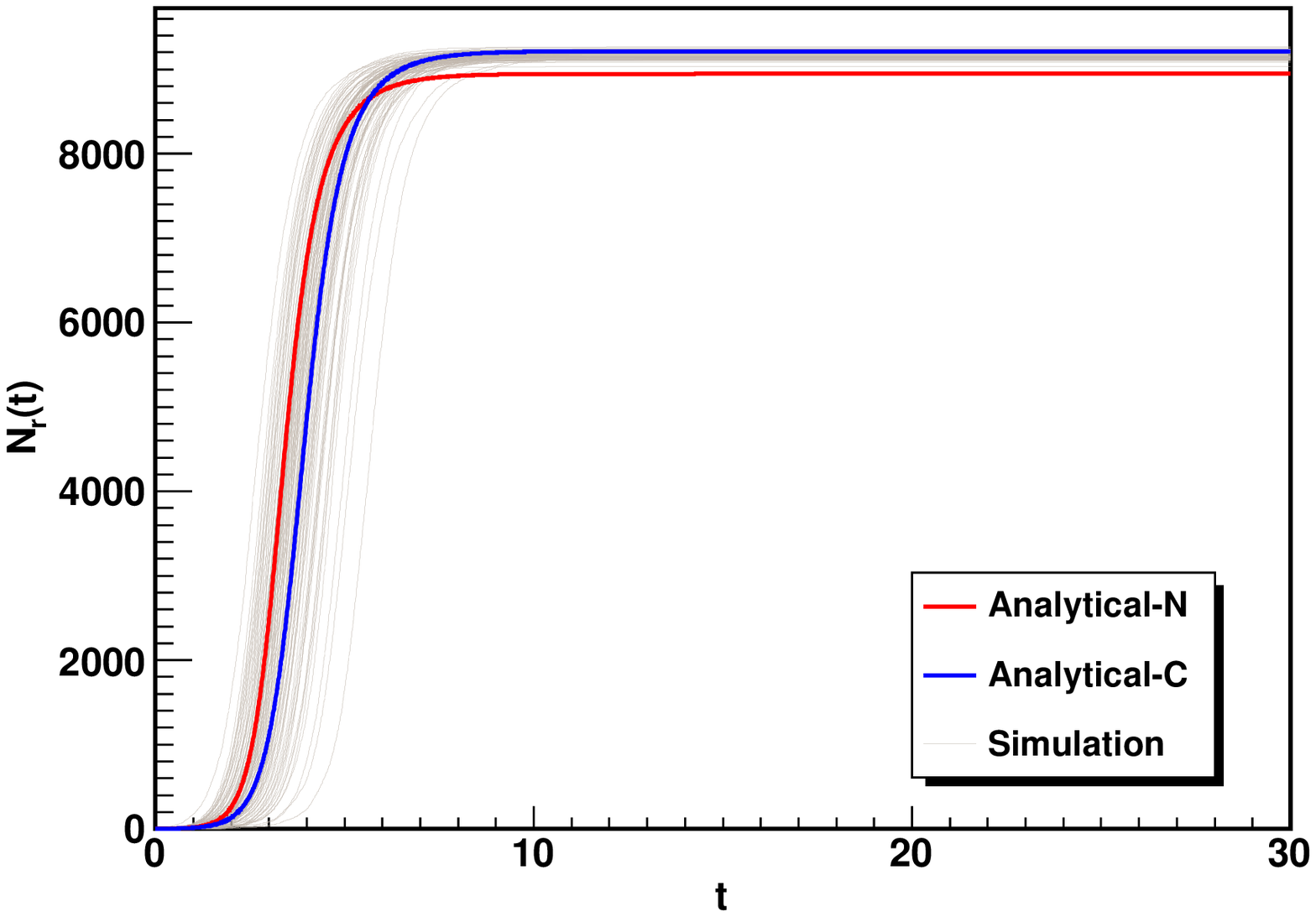}
\caption{$\lambda_{i}=0.25$ and $\lambda_{r}=2$ left top panel, $\lambda_{i}=0.2$
and $\lambda_{r}=1$ right top panel, $\lambda_{i}=0.5$ and $\lambda_{r}=2$
left bottom panel, and finally $\lambda_{i}=0.3$ and $\lambda_{r}=1$
right bottom panel.\label{fig:fig3}}
\end{centering}
\end{figure}

Finally in figure \ref{fig:fig3}, we compare the number of removed
vertices for the new calculation (Analytical-N) and compartmental
model (Analytical-C), versus simulations for bimodal network. Top
left panel $\lambda_{i}=0.25$ and $\lambda_{r}=2$; top right panel
$\lambda_{i}=0.2$ and $\lambda_{r}=1$; bottom left panel $\lambda_{i}=0.5$
and $\lambda_{r}=2$; and, bottom right panel $\lambda_{i}=0.3$ and
$\lambda_{r}=1$.

The small deviation for the current formalism and simulation for low
and high ${\cal R}_{0}$ values is due to overestimating/underestimating
the number of infections from the number of transmitting links. 

The compartmental model fails to correctly capture the dynamics of
the epidemics for populations in which a sizable portion of individuals
have low degree, which turns to be an important characteristic of
realistic human contact networks.

\section{Extensions}

In this section, we discuss different extensions of the methodology
to address other models such as, open system models (in which the
number of vertices is a function of time), dynamic network models,
susceptible-infectious-removed-susceptible SIRS models, and multi-type
systems (including systems with different age groups, each of which
having different connectivity, infectivity, susceptibility, \textit{etc}).

\subsection{Multi-type system}

Here we consider a system consisting of more than one type of nodes,where
each node type has different inter- and intra- degree distributions,
transmissibilities, and removal distributions. We define a set of
node types and index them with superscript $\alpha$. We first define
$N^{\alpha}$, $\lambda_{r}^{\alpha\beta}(\tau)$, $\lambda_{i}^{\alpha\beta}(\tau)$
and $p_{k}^{\alpha}$ as the number of vertices , new removal function,
infectivity function and degree distribution of type $\alpha$, respectively.
We also define $N_{\alpha}^{\beta}(t)$ as the number of vertices
from type $\beta$ in the classes $\alpha$. Here we assume the connectivity
between vertices, in different node types, is completely random. The
current method can be further extended if there is any preference
for intra- or inter-connectivity between vertices in different types.

Our renewal equation\eqref{eq:Jt} takes the following trivial form
\begin{equation}
J^{\alpha}(t)=\sum_{\beta}\int_{0}^{t}J^{\beta}(t-\tau)\Psi^{\beta}(\tau)\frac{dT^{\beta\alpha}(\tau,\tau<\tau_{r})}{d\tau}Z_{x}^{\beta\alpha}(t,\tau)d\tau\label{eq:Jtab}
\end{equation}

 $Z_{x}^{\beta\alpha}$, however, is now a more complex function and
must be calculated using the following steps; first, the collected
and uncollected degree distribution for each type must be found. To
do so we randomly pick a vertex. The probability that the first chosen
vertex is in type $\alpha$ is $n^{\alpha}(1)=N^{\alpha}/N$, as a
result the probability that the chosen vertex has degree $k$ and
comes from type $\alpha$ is given by $q_{k}^{\alpha}(1)=n^{\alpha}(1)p_{k}^{\alpha}$.
The degree distributions of collected and uncollected classes are
given by $\tilde{p}_{k}^{\alpha}(1)=q_{k}^{\alpha}(1)$ and $p_{k}^{\alpha}(1)=(N^{\alpha}p_{k}^{\alpha}-n^{\alpha}(1)\tilde{p}_{k}^{\alpha}(1))/(N-n^{\alpha}(1))$.
Now the second vertex can be randomly chosen by selecting a random
stub. The probability that the second vertex has degree $k$ and belongs
to type $\alpha$ is given by $q_{k}^{\alpha}(2)=n^{\alpha}(2)kp_{k}^{\alpha}(2)/z^{\alpha}(1)$
where $n^{\alpha}(2)=(N^{\alpha}-n^{\alpha}(1))/(N-1)$ and $z^{\alpha}(1)=\sum_{k}kp_{k}^{\alpha}(1)$.
The degree distribution of collected and uncollected classes are given
by $\tilde{p}_{k}^{\alpha}(2)=(q_{k}^{\alpha}(1)+q_{k}^{\alpha}(2))/(n^{\alpha}(1)+n^{\alpha}(2))$
and $p_{k}^{\alpha}(2)=(N^{\alpha}p_{k}^{\alpha}-(n^{\alpha}(1)+n^{\alpha}(2))\tilde{p}_{k}^{\alpha}(2))/(N-n^{\alpha}(1)-n^{\alpha}(2))$.
It is thus plausible that the $nth$ chosen vertex is in type $\alpha$
and has degree $k$ with the probability $q_{k}^{\alpha}(n)=n^{\alpha}(n)kp_{k}^{\alpha}(n-1)/z^{\alpha}(n-1)$.
In the same manner \begin{align}
\tilde{p}_{k}^{\alpha}(n) & =\frac{\sum_{i=1}^{n}q_{k}^{\alpha}(i)}{\sum_{i=1}^{n}n^{\alpha}(i)},\\
p_{k}^{\alpha}(n) & =\frac{N^{\alpha}p_{k}^{\alpha}-\sum_{i=1}^{n}n^{\alpha}(i)\tilde{p}_{k}^{\alpha}(n)}{N-\sum_{i=1}^{n}n^{\alpha}(i)}.\end{align}
 Therefore, $z_{j}^{\alpha}(t)$, $z_{r}^{\alpha}(t)$, $z_{i}^{\alpha}(t)$,
and $z_{s}^{\alpha}(t)$ are given by \begin{align}
z_{j}^{\alpha}(t) & =\tilde{z}^{\alpha}(N_{r}^{\alpha}(t)+N_{i}^{\alpha}(t))+[N_{r}^{\alpha}(t)+N_{i}^{\alpha}(t)]\left.\frac{d\tilde{z}^{\alpha}(n)}{dn}\right|_{n=N_{r}^{\alpha}(t)+N_{i}^{\alpha}(t)},\\
z_{r}^{\alpha}(t) & =\tilde{z}^{\alpha}(N_{r}^{\alpha}(t)),\\
z_{i}^{\alpha}(t) & =\frac{[N_{r}^{\alpha}(t)+N_{i}^{\alpha}(t)]\tilde{z}^{\alpha}(N_{r}^{\alpha}(t)+N_{i}^{\alpha}(t))-N_{r}^{\alpha}(t)\tilde{z}(N_{r}^{\alpha}(t))}{N_{i}^{\alpha}(t)},\\
z_{s}^{\alpha}(t) & =z^{\alpha}(N_{s}^{\alpha}(t)),\end{align}
 where $N_{i}^{\alpha}(t)=\int_{0}^{t}J^{\alpha}(t-\tau)\Psi^{\alpha}(\tau)d\tau$,
$N_{r}^{\alpha}(t)=\int_{0}^{t}J^{\alpha}(t-\tau)\left[1-\Psi^{\alpha}(\tau)\right]d\tau$
and $N_{s}^{\alpha}(t)=N-N_{i}^{\alpha}(t)-N_{r}^{\alpha}(t)$. Finally,
we have \begin{equation}
Z_{x}^{\alpha\beta}(t,\tau)\approx z_{j}^{\alpha}(t-\tau)\frac{z_{s}^{\beta}(t)N_{s}^{\beta}(t)}{z_{1}N}.\end{equation}
 As an example, we study an exponential degree distribution with $\kappa=10$,
$\lambda_{i}=0.2$ and $\lambda_{r}=1$. We use the multi-type framework
to show how easily one can replicate the results obtained earlier
for the similar degree distribution in section\{......\} the current
formalism. We devide the network to groups of vertices, each of which
has a specifc degree where $\{\alpha,\;\beta\}=\{k,\; k'\}$. Our
multi-type renewal equations is given by \begin{equation}
J^{\alpha}(t)=\int_{0}^{t}\Psi(\tau)\frac{dT(\tau,\tau<\tau_{r})}{d\tau}\sum_{\beta}J^{\beta}(t-\tau)Z_{x}^{\beta\alpha}\frac{N_{s}^{\alpha}(t)}{N^{\alpha}}d\tau\label{eq:J2C}\end{equation}
 A crude approximation for the contact matrix, ${\cal Z}_{x}$, is
described below. The probability of a stub from a node type $\alpha$
connecting to a stub from node type $\beta$ is given by $N^{\beta}z_{1}^{\beta}/Nz_{1}$.
This implies that the total number of links going from type $\alpha$
to $\beta$ can be approximated by $N^{\alpha}z_{1}^{\alpha}N^{\beta}z_{1}^{\beta}/Nz_{1}$
and consequently the number links per vertex is given by $z_{1}^{\alpha}N^{\beta}z_{1}^{\beta}/Nz_{1}$.

\begin{figure}[htb]
\begin{centering}
\includegraphics[scale=0.5]{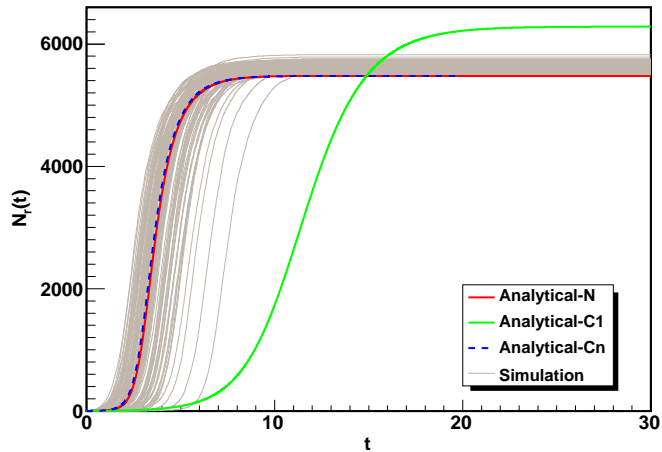}
\caption{$\kappa=10$, $\lambda_{i}=0.2$ and $\lambda_{r}=1$.
\label{fig:fig4}}
\end{centering}
\end{figure}

In figure \ref{fig:fig4} we compare the result of the current calculation
(Analytical-N) against one type (Analytical-C1), multi-type (Analytical-Cn)
and simulation models for the above-mentioned network, whereby $\lambda_{i}=0.2$
and $\lambda_{r}=1$. The excellent agreement between the two methods
demonstrates that a network with a general degree distribution can
be examined as a set multi-type system, within the current approximation.
Both the multi-type framework and the current formalism, as discussed
in section\ref{sec:Num_simp_net}, yield similar levels of error when
predicting the epidemic curve. Thus, one can use either approach;
however, the multitype approach may become very expensive computationally
for a network with a very wide degree distribution. %

\subsection{Open system}

In an open system, the number of vertices is a function of time. The
previous set of equations still hold for an open system, however,
we must now keep track of entering and exiting vertices in each class,
as well as the corresponding change in degree distribution. For example,
the number of vertices in a susceptible class can be calculated from
\begin{equation}
\frac{dN_{s}(t)}{dt}=\sum_{k}(\Pi_{in,s}(k,t)-\Pi_{out,s}(k,t)),\end{equation}
 where $\Pi_{in/out,s}(k,t)$ is the rate of entry/exit of susceptible
vertices with degree $k$ at time $t$. The degree distribution of
collected and uncollected vertices can be calculated from \begin{align}
\frac{d\tilde{p}_{k}(n,t)}{dt} & =\frac{\partial\tilde{p}_{k}(n,t)}{\partial n}\frac{dn}{dt}+\frac{\Pi_{in,ir}(k,t)-\Pi_{out,ir}(k,t)}{n(t)},\label{eq:dpdto}\\
\frac{dp_{k}(n,t)}{dt} & =\frac{\partial p_{k}(n,t)}{\partial n}\frac{dn}{dt}+\frac{\Pi_{in,s}(k,t)-\Pi_{out,s}(k,t)}{N(t)-n(t)},\label{eq:dptdto}\end{align}
 where $\Pi_{in,ir}(k,t)=\Pi_{in,i}(k,t)=\Pi_{in,r}(k,t)$. The first
term on the right hand side of both equations is the contribution
of collecting vertices, the second term arises from vertices entering
or exiting the network. The partial derivatives of both equations
are calculated from \eqref{eq:pktc} and \eqref{eq:pkc} respectively
\begin{align}
\frac{\partial\tilde{p}_{k}(n,t)}{\partial t} & =\frac{p_{k}(n(t),t)}{n(t)}\left(\frac{k}{z(n(t))}-1\right),\\
\frac{\partial p_{k}(n,t)}{\partial t} & =\frac{p_{k}(n(t),t)}{N(t)-n(t)}\left(1-\frac{k}{z(n(t))}\right).\end{align}

\subsection{Dynamic network}

As another possible extension, we consider networks where the degree
of each vertex is a function of time. Dynamic networks are a simple
example of an open system with the constrains \begin{equation}
\sum_{k}\Pi_{in,\alpha}(k,t)=\sum_{k}\Pi_{out,\alpha}(k,t)\end{equation}
 where $\alpha=\{s,\; i,\; r\}$ is an index for susceptible, infectious
and removed vertices. Accordingly, the outflow of vertices with a
specific degree from a given class should be replaced by the same
number of vertices but with different degrees. This is a consequent
of the fact that infection is instantaneous and that a removed vertex
remains removed. $\Pi_{in/out,\alpha}(k,t)$ could have complex dynamics
as long as the above constraints are satisfied.

\subsection{SIRS model}

For the SIRS model, we first need to introduce the probability function,
which specifies the chance of re-infection over time, once the infected
vertex has recovered. This variable is generally a function of time
and can be measured with respect to any infection time reference.
We define the susceptibility function, $\lambda_{s}(\tau)$, in which
$\lambda_{s}(\tau)d\tau$ gives the probability of an infected vertex
becoming susceptible again in the interval $\tau$ and $\tau+d\tau$.
We define $\Psi_{\alpha\beta}(\tau_{s})$ as a probability function
which give the probability of the time of movement from disease state
$\alpha$ to $\beta$, moreover \begin{equation}
\Psi_{\alpha\beta}(\tau_{s})=\exp\left(-\int_{0}^{\tau_{r}}\lambda_{\alpha\beta}(u)du\right)\end{equation}
 The number of susceptible and removed vertices is given by \begin{align}
N_{s}(t) & =N-\int_{0}^{t}J(t-\tau)d\tau+\int_{0}^{t}J(t-\tau)[1-\Psi_{rs}(\tau)]d\tau,\\
N_{r}(t) & =\int_{0}^{t}J(t-\tau)[1-\Psi_{ir}(\tau)]d\tau-\int_{0}^{t}J(t-\tau)[1-\Psi_{rs}(\tau)]d\tau\nonumber \\
 & =\int_{0}^{t}J(t-\tau)[\Psi_{rs}(\tau)-\Psi_{ir}(\tau)]d\tau.\end{align}
 The degree distribution of collected and uncollected vertices is
calculated as follows \begin{align}
\frac{d\tilde{p}_{k}(n,t)}{dt} & =\frac{\partial\tilde{p}_{k}(n,t)}{\partial n}\frac{dn}{dt}-\frac{\Pi_{out,r}(k,t)}{n(t)},\label{eq:dpdt}\\
\frac{dp_{k}(n,t)}{dt} & =\frac{\partial p_{k}(n,t)}{\partial n}\frac{dn}{dt}+\frac{\Pi_{in,s}(k,t)}{N(t)-n(t)},\label{eq:dptdt}\end{align}
 where $\Pi_{out,r}(k,t)$ is the outflow of recovered vertices to
susceptible classes and in the same manner $\Pi_{in,s}(k,t)$ is inflow
of susceptible vertices from the recovered class; this calculation
involves defining the rate of outgoing recovered vertices as \begin{equation}
J_{out}(t)=\frac{d}{dt}\int_{0}^{t}J(t-\tau)(1-\Psi_{rs}(\tau))d\tau.\end{equation}
 We also define the rate of outgoing degree of recovered vertices
as \begin{equation}
z_{out}(t)=\frac{d}{dt}\int_{0}^{t}J(t-\tau)z_{j}(t-\tau)(1-\Psi_{rs}(\tau))d\tau.\end{equation}
 The expected degree of outgoing recovered vertices is then given
by \begin{equation}
k_{out}(t)=\frac{\zeta_{out}(t)}{J_{out}(t)},\end{equation}
 and as a result \begin{equation}
\Pi_{out,r}(k,t)=\Pi_{in,s}(k,t)=\delta_{k,k_{out}(t)}J_{out}(t).\end{equation}
 In practice $k_{out}(t)$ is not an integer function since it gives
the average degree of new susceptible vertices; thus, one must properly
distribute new vertices around $k_{out}(t)$ to ensure that the average
degree of the system remains constant.

\section{Conclusions and Discussion}

The novel methodology outlined above allows us to evaluate the time
evolution of disease spread on a network. Our methodology is able
to accommodate diseases with very general infectivity profiles. Additionally,
this methodology can manage multi-type networks, dynamical networks,
and SIRS systems. The precision of this methodology depends on the
accuracy of the kernel of the renewal equation for the infection rate,
which will be the subject of future investigations.

\section{Acknoledgment}

BP would like to acknowledge the support of the Canadian Institutes
of Health Research (grant nos. MOP-81273, PPR-79231 and PTL-97126
{[}Team Leader grant (CanPan II){]}) and the Michael Smith Foundation
for Health Research (Senior Scholar Funds). BD was supported by these
grants.

\end{document}